\documentclass[10pt]{article}
\usepackage{amsmath}
\usepackage{amsfonts}
\usepackage{epsfig}
\usepackage{color}

\input epsf

\begin{document}

\begin{center}
{\LARGE Analytic models for mechanotransduction: gating a mechanosensitive channel} \\
\vspace{0.5cm}
\textsc{ Paul Wiggins${}^*$ and  Rob Phillips${}^\dag$ }\\
\vspace{0.25cm}

${}^*$Division of Physics, Mathematics \& Astronomy, \\
${}^\dag$Division of Engineering and Applied Science, \\
California Institute of Technology, \\
1200 E. California Blvd. \\ Pasadena, California 91125-9500. 

Correspondence should be addressed to R.P. email: phillips@aero.caltech.edu 

\end{center}

\begin{abstract}
\textbf{
Analytic estimates for the forces and free energy generated by bilayer
deformation reveal a compelling and intuitive model for MscL channel
gating analogous to the nucleation of a second phase. We argue that
the competition between hydrophobic mismatch and tension
results in a surprisingly rich story which can provide both
a quantitative comparison to 
measurements of opening tension for MscL when reconstituted in 
bilayers
of different thickness and qualitative insights into the function of
the MscL channel and other transmembrane proteins.
}
\end{abstract}


\section{Introduction}
The mechanosensitive channel (MscL) is a compelling example of the interaction between a protein and 
the surrounding bilayer membrane. The channel is gated mechanically by applied tension and
is believed to function as an emergency relief valve in bacteria \cite{moe}.
MscL is a member of  a growing class of proteins which have been determined to be mechanosensitive 
\cite{Gill}, \cite{hamill}. 
The dependence of the conductance on applied tension has been studied extensively in patch clamp 
experiments \cite{suk1}, \cite{perozo1}, \cite{perozo2}. 
In terms of the observed conductance, these studies have revealed that the channel is very 
nearly a two state system. MscL spends the vast majority of its life in either a closed state (C) 
or an open state (O) characterized by a discrete conductance.
When the bilayer tension is small, the protein is exclusively in the closed configuration. 
As the tension grows, the open state 
becomes ever more prevalent, until it dominates at high tension. The simplest structural 
interpretation of this conductance data is to assume that each discrete conductance corresponds 
to a well defined channel conformation. This assumption seems to be compatible with the conductance 
data\footnote{The area of the states seems roughly independent of applied tension \cite{suk1}.}.
Patch clamp experiments have also revealed that there are at least three additional 
discrete, intermediate conductance levels \cite{suk1}
suggesting three additional short lived substates (S1-S3). 
Rees and coworkers \cite{Rees} have solved the structure for one conformation using X-ray 
crystallography. This state appears to be the closed state \cite{Rees}, \cite{perozo2}.  
Perozo {\it et al.} \cite{perozo2} have trapped MscL in the open 
configuration and used electron paramagnetic resonance spectroscopy (EPR) and site-directed spin 
labeling (SDSL) to deduce its geometry. Sukharev {\it et al.} \cite{suk2} 
have also proposed an open state conformation based on structural considerations. 

The conformational landscape of the MscL channel is extremely complex, depending on huge number
of microscopic degrees of freedom which are analytically intractable. Even from the standpoint of 
numerical calculations, this number is still very large \cite{md1}. As an alternative
to a detailed microscopic picture of MscL, we consider a simplified free energy function where we 
divide the free energy of the system into two contributions, namely,
\begin{equation}
G = G_P+G_{\cal M},
\end{equation}
where $G_{P}$ is the free energy associated with the conformation of the protein and $G_{{\cal M}}$ is the 
deformation free energy from the bulk of the bilayer \cite{huang}. In general, these two terms are 
coupled. The conformation of the protein depends on the forces applied by the bilayer. The bilayer 
deformation is induced by the external geometry of the protein. We denote this external geometry with a state 
vector, $X$, which captures the radius of the channel as well as its orientation relative to the surrounding
bilayer as described in more detail below. We calculate the induced bilayer deformation energy, $G_{\cal M}(X)$,
by minimizing the free energy of the bilayer and solving the resulting boundary value problem
using an analytic model developed for the study of bilayer mechanics \cite{helfrich} and protein-bilayer 
interactions \cite{huang}, \cite{nielsen}, \cite{goulian}, \cite{ndan4}. 
We then apply asymptotic approximations to the exact solutions 
of this model for cylindrically symmetric inclusions, permitting all of the results to be expressed, 
estimated, and understood with simple scaling relations. 
The advantage of this model is that it permits us to characterize the protein-bilayer system
in a way that is at once analytically tractable and predictive. 
By understanding the consequences of the simplest models, we develop a framework in which to understand 
the richer dynamics of the real system. There is a wealth of useful, physical intuition to 
be gleaned from this model relating to both the function of MscL and 
that of mechanosensitive transmembrane proteins more generally. In a forthcoming paper, 
we will show that the mechanics of the bilayer must play an integral role in  
mechanotransduction and channel function. Specifically, we will present detailed analytic estimates of the
free energy generated by bilayer deformation induced by the channel and show that these free energies 
are of the same order as the free energy differences measured by Sukharev {\it et al.} \cite{suk1}.
These analytic calculations reveal a compelling and intuitive model for the gating of the MscL channel 
which is the subject of this current paper. The competition between hydrophobic mismatch and applied 
tension, in the presence of radial constraints, generates a bistable system that is implicitly 
a mechanosensitive channel. Furthermore, this simple model provides a picture
which is both qualitatively and quantitatively consonant with the measured  
 dependence of the free energy on acyl chain length as observed by Perozo {\it et al.} \cite{perozo1}. 
In addition, these results may also explain the stabilization of the open state by spontaneous 
curvature inducing lysophospholipids observed by Perozo {\it et al.} \cite{perozo1}, although more experiments are 
required to check the consistency of this proposal.

\begin{figure}
\begin{center}
\leavevmode 
\epsfig{file=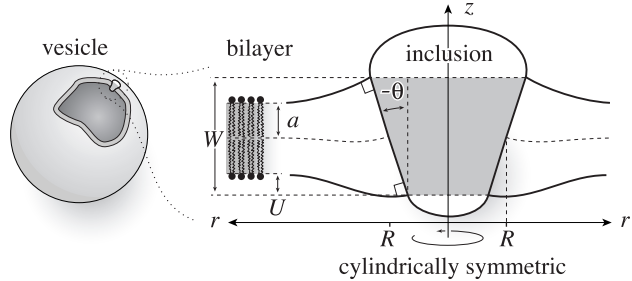}
\caption{\label{basic} The bilayer-inclusion model. The geometry of the inclusion 
is described by three parameters: the radius, $R$, the hydrophobic thickness, $W$, and the radial mid-plane 
slope, $H'$. The hydrophobic mismatch, $2U$, is the difference between the hydrophobic protein thickness, $W$, 
and the bilayer equilibrium thickness, $2a$. We assume the surfaces of the bilayer are locally normal to 
the interface of the inclusion, as depicted above, implying that the mid-plane slope is related to the 
interface angle:  $H'=\tan \theta$.}
\end{center}
\end{figure}

\section{The energy landscape of the bilayer}
\label{ELB}
\begin{figure} 
\begin{center}
\epsfig{file=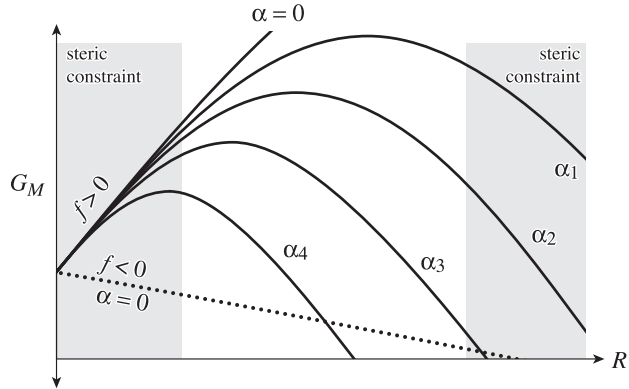}
\caption{\label{FMR} The bilayer deformation energy landscape. The bilayer deformation
energy is plotted as a function of the radius
for different values of applied tension. The solid curves represent the bilayer deformation 
energy with a positive line tension, $f$, 
for various different tensions ($0<\alpha_1<\alpha_2<\alpha_3<\alpha_4$). 
The competition between interface energy and applied tension naturally gives rise to a 
bistable potential when the radial domain is limited by steric constraints. 
The gray regions represent radii inaccessible to the channel due to steric constraints. 
These constraints are briefly motivated in section \ref{ELB}. If the line tension
is negative, depicted by the dotted curve, the potential is never bistable.}
\end{center}
\end{figure}
In the calculations considered here, the geometry of the protein, characterized by
the conformational state vector $X$,  is described by three geometrical parameters
\begin{equation}
X = (R, W, H'), 
\end{equation}
where $R$ is the radius of the channel, $W$ is the hydrophobic thickness, and 
$H'$ is the mid-plane slope. See figure \ref{basic} for details.
\label{ELB}
Although we have parameterized the conformation space of the protein with these three parameters, in this paper,
we will focus on the radial dependence alone, claiming that even in this reduced description, 
the model provides a rich variety of predictions which are compatible with previous observation and suggest new
experiments.  
The radial dependence of the bilayer deformation energy is particularly 
important for MscL since the radius undergoes a very large change between the open and closed states \cite{suk2}.
The bilayer deformation energy can be written explicitly in terms of the channel radius as
\begin{equation}
\label{bde}
G_{\cal M}=G_0+f\cdot 2\pi R-\alpha\cdot\pi R^2,
\end{equation}
where $G_0$ and $f$ do not explicitly depend on $R$ and $\alpha$ is the applied tension which triggers 
channel gating.
$G_0$ is a radially independent contribution to the deformation energy which is a function of the other 
geometrical parameters of the protein. Its importance in gating the channel is most likely secondary since 
it is independent of $R$ and it will be ignored in the remainder of the discussion.
The dependence of bilayer deformation energy on applied tension can be explained intuitively  \cite{hamill}. 
The free energy contribution for a small change in the channel area due to the applied tension can be written 
$-\alpha dA$ which is the two dimensional analogue of the $-PdV$ term for a gas in three dimensions. 
At high enough applied tension, the state with the largest inclusion area will have the lowest free energy. 

The line tension, $f$, contributes an energy proportional to the circumference. 
Line tensions arise very naturally since the interface area of the inclusion is proportional 
to the circumference.  In what follows, we will discuss the two dominant contributions to this line tension: 
thickness deformation \cite{huang}, \cite{nielsen}, \cite{goulian} and spontaneous curvature 
\cite{ndan4}. The thickness deformation free energy is induced by the mismatch between the 
the equilibrium thickness of the bilayer and the hydrophobic thickness of the protein. 
The importance of this hydrophobic mismatch in the function of transmembrane proteins has 
already been established \cite{Killian}.
The bilayer deforms locally to reduce the mismatch with the protein as shown in figure~\ref{basic}. 
Symbolically the thickness deformation energy is \cite{huang},
\begin{equation}
G_U = f_U\cdot 2\pi R= {\textstyle \frac{1}{2}} {\cal K} U^2 \cdot 2\pi R,
\end{equation}
where ${\cal K}=2\times 10^{-2}\ kT\ {\rm \AA}^{-3}$ is an effective elastic modulus defined in 
the appendix which is roughly independent of acyl chain length\footnote{See the appendix for 
an brief discussion of scaling of the elastic constants.} and $U$ is half the hydrophobic mismatch as defined 
in figure \ref{basic}. 
Naturally the energetic penalty for this deformation is 
 proportional to the mismatch squared since the minimum energy
 state corresponds to zero mismatch. The area of that part of the bilayer
 which is  deformed  is roughly equal to  the circumference of the channel times an elastic 
decay length.  As a result, the contribution of thickness
deformation to the total free energy budget scales with the radial dimension
of the channel.   We also note that the thickness deformation
free energy is always positive. 

In contrast, the free energy induced by spontaneous curvature 
can be either negative or positive. Physically, this free energy comes from locally relieving or increasing 
the curvature stress generated by lipids or surfactants which induce spontaneous 
curvature \cite{ndan4}, \cite{israelachvili}, \cite{gruner}. Again the radial dependence of this free energy will be linear since 
the effect is localized around the interface. Since the leaflets of the bilayer can be doped 
independently \cite{perozo1}, the spontaneous curvatures of the top and bottom leaflets, $C_\pm$,  
can be different. It is convenient to work in terms of the composite spontaneous curvature of the bilayer, 
$C \equiv \textstyle \frac{1}{2}(C_+-C_-)$.
The contribution to the deformation energy arising from spontaneous curvature is given by \cite{ndan4}
\begin{equation}
G_C = f_C\cdot 2\pi R = K_B C H'\cdot 2\pi R,
\end{equation}
where $H'$ is the mid-plane slope and $K_B = 20(a/20\ {\rm \AA})^3\ kT$,
is the bending modulus which roughly scales as the third power of the bilayer thickness\footnote{See the appendix for 
an brief discussion of scaling of the elastic constants.}.
We will discuss these results in more detail elsewhere.
Notice that if $C$ and $H'$ have opposite signs, the deformation 
energy and the corresponding line tension, $f_C$, will be negative. 
We note that the elastic theory of membrane deformations associated with proteins like MscL permit
other terms (such as mid-plane deformation, for example) which can be treated within the same framework
and which give rise to the same radial dependence as that described here.  However, for
the purposes of characterizing the energetics of MscL, these other terms are
less important than the two considered here.

Typically, in the absence of large spontaneous curvature, the line tension,
$f$, will be dominated by the mismatch and will be positive. A potential of the form described by equation 
\ref{bde} is depicted schematically in figure \ref{FMR}.  
In this figure, we have implied that there are steric constraints for the range of radii accessible to the 
protein. Assuming that there is a lower bound on the radius of the inclusion is very natural. It can 
be understood as the radius below which the residues begin to overlap. This steric constraint will generate
a hard wall in the protein conformation energy, forbidding lower radii. Similar but slightly more elaborate 
arguments can made for an upper bound. The bilayer deformation energy generates a barrier between small radius 
and large radius states. The location of the peak of this barrier 
is the turning point
\begin{equation}
R_* \equiv \frac{f}{\alpha}.
\end{equation} 
At small tension, the turning point is very large and is irrelevant since it occurs at a radius not 
attainable by the channel due to the steric constraints, but as the tension increases the position of the
turning point decreases. This behavior is reminiscent of the competition between surface tension and 
energy density for nucleation processes which give rise to a similar barrier ({\it e.g.} \cite{kittel}). 

\subsection{Induced tension and bilayer induced stabilization}
\begin{figure} 
\begin{center}
\epsfig{file=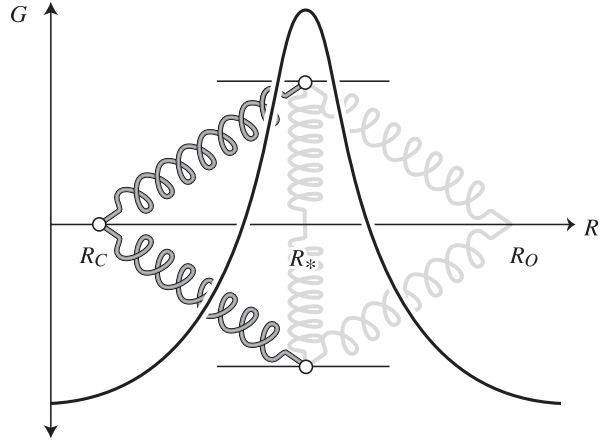}
\caption{ \label{mech} 
The free energy of a bistable system. In the background of the plot, we have 
depicted a mechanical system that is bistable. At the equilibrium length of the
springs, the hinge can be at either $R_C$ or $R_O$. To move the hinge between 
these two stable states it must be compressed, giving rise to a barrier and 
a transition state at $R_*$. This potential is analogous to that generated by 
bilayer deformation. This picture also serves as a schematic potential for 
the gating transitions as described in the text.
}
\end{center}
\end{figure}
\begin{figure} 
\begin{center}
\epsfig{file=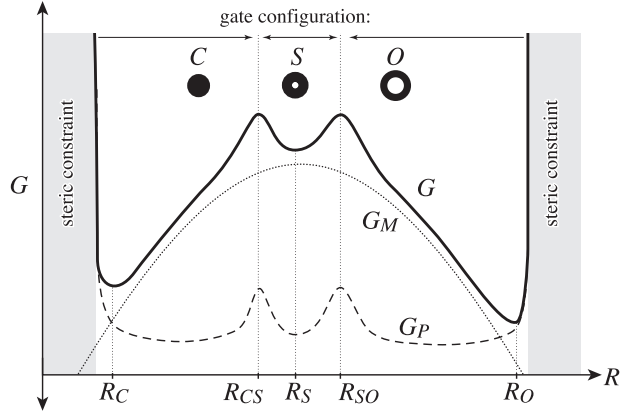}
\caption{ \label{FMR2} 
The total free energy. The total free energy, $G$, of the protein and bilayer are 
plotted schematically as a function of channel
radius. The bilayer deformation energy, $G_{\cal M}$, is represented by the dotted 
curve. A schematic protein conformation energy is represented by the dashed curve. 
Their sum gives the total free energy $G$. The protein energy has been chosen to
contain a single substate, $S$. There is a conformational energy barrier corresponding 
to changing the gate conformation of the channel. These transitions occur at $R_{CS}$
and $R_{SO}$. $G_P$ also contains steep barriers corresponding to steric 
constraints. The radii of the conductance states are defined by the free energy minima of $G$.
As we have discussed in the text, the 
bilayer deformation energy can stabilize either or both the closed and open states 
since they are the states of highest and lowest radius. At the tension depicted above,
both have been stabilized. In contrast, the substates are stabilized by protein conformation 
but never by the bilayer.  }
\end{center}
\end{figure}

\label{defforce}
Although the conformational landscape of the MscL channel is certainly very complicated,
there is an intriguing possibility that the channel harnesses the elastic properties of the bilayer 
which quite naturally provide the properties we desire in a mechanosensitive channel: 
a stable closed state at low tension
and a stable open state at high tension. 
In effect, we will treat the bilayer deformation energy as an external 
potential with respect to the conformational energy landscape of the protein. 
The physical effects of the radial dependence of the bilayer deformation energy 
on the inclusion conformation can be recast in a more intuitive form by appealing 
to the induced tension which accounts not only for the
applied far field tension, but also for {\it induced} tension
terms due to bilayer deformation. The applied tension is not the whole story!  
The generalized forces are obtained by differentiating the bilayer deformation energy 
with respect to bilayer excursions. The net tension induced by the bilayer on the 
inclusion interface is
\begin{equation}
\alpha_\Sigma = \alpha - \frac{f}{R},
\end{equation}
where we have denoted the net tension $\alpha_\Sigma$ since we have already used $\alpha$ 
to denote the applied tension. For radii smaller than the turning point, $R_*$, the 
bilayer deformation energy is an increasing function of radius and therefore the net 
tension is negative and acts to compress the channel. For radii larger than that at the turning point, 
the bilayer deformation energy is a decreasing function of radius and the net tension is positive 
and acts to expand the channel. 
The combination of these constraints and the bilayer deformation energy lead to a bistable system where
the closed and open states correspond to the constraint-induced radial minimum
and maximum, respectively. Recall that the net tension on the closed state will be compressive as long 
as its radius is smaller than that at the turning point, namely,
\begin{equation}
R_C<R_*=\frac{f}{\alpha}.
\end{equation}
This inequality defines the range of applied tension over which the closed state is stabilized by the 
bilayer deformation energy. The net tension on the open state will be expansive as long as its
radius is greater  than that at the turning point:
\begin{equation}
R_O>R_*=\frac{f}{\alpha}.
\end{equation}     
This inequality defines the range of applied tension over which the open state is stabilized by the 
bilayer deformation energy. There is an intermediate range of tensions for which both states are 
stabilized by the bilayer,
\begin{equation}
\frac{f}{R_O} < \alpha < \frac{f}{R_C}.
\end{equation} 
The bilayer deformation energy naturally destabilizes the open state for applied tension 
below this range while stabilizing the closed state for applied tensions up to the limit of this range.  
Both effects help to prevent the channel from leaking at low applied tension. This bistability is precisely
the desirable behavior for a mechanosensitive channel designed to relieve internal pressure and yet 
surprisingly little is required from the protein conformational landscape, $G_P$, except for steric 
constraints which arise very naturally. 
In figure \ref{mech}, 
we have depicted a mechanical analogue of this bistable system. In figure \ref{FMR2}, we have depicted the
way in which MscL mimics this mechanical analogue by using the sum of a schematic protein energy and the 
bilayer deformation energy to form energy minima corresponding to the open and closed states. 

Our general discussion of the role of the bilayer enables us to make some rather 
general observations about the nature of the substates. In order to generate a substate,
we assume that there is more than one gating transition in the protein conformational energy. 
One gating transition would correspond to a closed to open transition. An additional gating transition allows
three conductance states. We will assume that these transitions are themselves bistable in 
nature since the conductance data would seem to imply the lifetimes of the transition states
are very short compared to the conductance states \cite{suk1}.
In other words, the conformational gating transition occurs near a local maximum in the 
conformational free energy like that depicted in figure \ref{mech}. If we add two such 
transitions to $G_P$, we generate a substate of intermediate radius between these two transition 
state radii. A schematic example of this is illustrated in figure \ref{FMR2}.
Sukharev {\it et al.} \cite{suk1} have shown that all the substates are short lived and have estimated the 
areas of each state based on the tension dependence of their free energies\footnote{There is now evidence for 
additional substates \cite{suk3}.}. Specifically, they
have shown that the radii of the substates lie between the open and closed state radii.
If the bilayer deformation dominates the free energy of the states, 
the ephemeral nature of the substates is a natural consequence of their intermediate radii.
The compressive tension due to the mismatch stabilizes the state of lowest radius at low applied tension. At high 
applied tension the bilayer stabilizes the state with highest radius. All the states with 
intermediate radii are never stabilized by the bilayer and are therefore short lived.
Our deceptively simple mismatch model quite naturally leads to short lived substates 
at intermediate radii.  

In this section we have shown that some very general assumptions about the form of the 
bilayer deformation energy lead to  channel behavior which is mechanosensitive. 
Although we have suggested a specific source, thickness deformation, for the line tension, $f$, 
this linear radial dependence arises very naturally. For instance, if we assume the 
bilayer is too stiff to deform and the mismatch causes hydrophobic residues to come
into contact with the solvent, a larger positive line tension would be induced.
The competition between line tension and applied tension is therefore a quite general 
phenomena for mechanosensitive transmembrane proteins.  We shall demonstrate below that, for MscL, 
this mechanism appears to explain the free energy dependence on bilayer acyl chain length 
surprisingly well.      

\section{ Results}
\label{mscl_g}
The patch clamp experiments of Perozo {\it et al.} \cite{perozo1} go beyond the 
earlier work of Sukharev {\it et al.} \cite{suk1}  by providing experimental values 
for the free energy difference between the open and closed states for bilayers of 
several thicknesses. These results can be compared with our predictions. In order to
apply our model, we must determine the geometrical parameters of the state vector $X$
for the open and closed states and in particular the open and closed radii. 
The radius of the closed state is known from x-ray 
crystallography \cite{Rees}: $R_C\sim 23\ \rm \AA$. Structural studies \cite{suk2} and 
EPR and SDSL \cite{perozo2} experiments have suggested an open state radius of roughly
$R_O\sim 35\ \rm \AA$. In order to estimate the line tension and free energy generated by 
hydrophobic mismatch, we must determine the hydrophobic thickness $W$. (We  ignore the 
difference  in hydrophobic thickness between the open and closed states.) In principle, 
one might have thought this could be deduced from the atomic-level structure of MscL, 
but in practice, real structures are complicated, often 
lacking a clear transition from hydrophobic to hydrophilic residues on the interface.
However, this width may be deduced from the EPR and SDSL data of Perozo {\it et al.} \cite{perozo1}. 
EPR and SDSL experiments measure inter-subunit proximity 
and spin-label mobility, respectively \cite{perozo1}. Compressive tension in the bilayer 
suppresses the fluctuations of the protein, increasing the subunit proximity and reducing the 
spin label mobility. In the experiments of Perozo {\it et al.} \cite{perozo1}, 
the applied tension is low implying that  
the net tension is dominated by the line tension, induced by thickness deformation,
\begin{equation}
\label{induced_tension}
\alpha_{\Sigma} \sim -\frac{{\cal K} U^2}{2R}, 
\end{equation} 
in the absence of spontaneous curvature.
This tension is compressive and proportional to the mismatch squared. Therefore, when the mismatch
is zero, the tension reaches a minimum, implying that mobility and subunit separation should reach a 
maximum. The EPR and SDSL data of Perozo {\it et al.} \cite{perozo1}
may turn over for PC12 bilayers, implying that the mismatch is zero which would imply in turn that
\begin{equation}
W \sim 2a_{n=12}, 
\end{equation} 
but due to the quadratic dependence on $U$, the slope in the vicinity of the turnover is small.  
Since PC lipids with acyl chain length shorter than $n=10$ do not form stable bilayers \cite{perozo1}, 
it is difficult to extensively check the quadratic dependence on $U$. The predicted
turnover would be more pronounced for PC bilayers with $n<10$. We shall see that 
this deduced hydrophobic mismatch is compatible with the patch clamp measurements of 
Perozo {\it et al.} \cite{perozo1}.

\subsection{Hydrophobic mismatch}
\label{hm}
\begin{figure}
\begin{center}
\epsfig{file=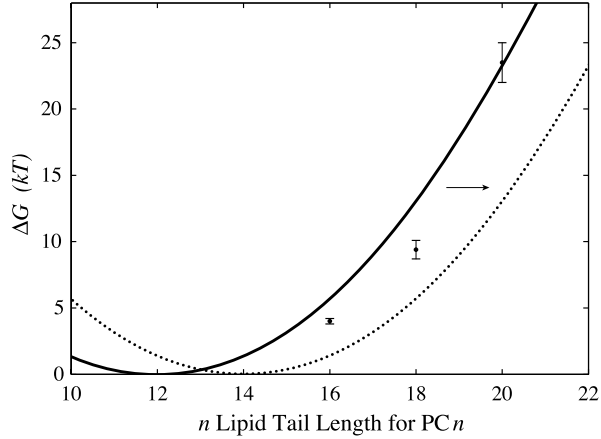}
\caption{\label{res} Free energy difference between open
and closed states vs lipid acyl chain length. 
The experimental data of Perozo {\it et al.} \cite{perozo1} for the free energy 
difference between the open and closed states at zero tension, $\Delta G_0$, is plotted 
with black circles and error bars. The solid curve represents the theoretical values 
for the bilayer deformation energy generated by a simple thickness deformation model at 
zero tension, $\Delta G_{0,\cal M}$.
 This correspondence is strong evidence for a central role
for the hydrophobic mismatch in the function of MscL. The dotted curve represents 
the translated $\Delta G_{0,\cal M}$ for an engineered MscL channel with a hydrophobic 
thickness matching a PC14 bilayer.
}
\end{center}
\end{figure}    
Perozo {\it et al.} \cite{perozo1} have measured $\Delta G_0$, the free energy difference 
between the open and closed state at zero tension\footnote{The free energy measured by Perozo {\it et al.} 
\cite{perozo1} is equivalent to the free energy at zero tension modulo several assumptions \cite{hamill}.}
for three acyl chain lengths. 
Using the value we have deduced for $W$, we can now calculate the free energy difference between the
open and closed states due to bilayer deformation at zero tension, $\Delta G_{0,\cal M}$, which is given by 
the line tension contribution alone:
\begin{equation}
\Delta G_{0,\cal M} = f2\pi\Delta R \rightarrow f_U2\pi \Delta R,
\end{equation}
where $\Delta R$ is the difference between the open and closed radii.  
The theoretical result, $\Delta G_{0,\cal M}$, is plotted with the experimental measurements of
$\Delta G_0$ in fig \ref{res}. The 
agreement between experiment and theory is embarrassingly good given the cavalier fashion 
in which we have chosen the geometrical parameters and that we have neglected the protein conformational energy, $G_P$,  
entirely. 
There is a very important point to be made about these results. Perozo {\it et al.} \cite{perozo1}
have measured three data 
points and our model is quadratic, 
implying that we could have chosen the parameters of our model to fit the data 
points perfectly since any three points lie on a parabola, but our parameters have in fact been deduced
independently rather than fit, which is why 
this correspondence with the data is remarkable. This model corresponds to a channel where the free 
energy difference between the open and closed states is dominated by the bilayer deformation
rather than protein conformation. 
Our model implies that $\Delta G_0$ for PC10, PC12, and PC14 should be very small. Unfortunately these 
bilayers have proved too weak for patch-clamp measurements of $\Delta G_0$ \cite{perozo1}. 
Certainly none of these bilayers trap the channel in the open state \cite{perozo1}. 

The opening tension is defined as the tension at which the open and closed state probabilities 
are equal or, analogously, the tension at which the free energies of the open and closed states are equal. 
The opening tension is
\begin{equation}
\label{opening_tension}
\alpha_{1/2} = \frac{f}{\overline R}+\frac{\Delta G_P}{\Delta A},
\end{equation}
where ${\overline R}\equiv {\textstyle \frac{1}{2}}(R_C+R_O)$ is the mean radius, $\Delta G_P$ is the difference in 
the open and closed state protein conformation energy, and $\Delta A$ is the difference in open
and closed state area. When the bilayer deformation energy dominates, the opening tension is determined by the 
first term alone. Changing the sensitivity of the channel is straightforward from 
this perspective. Changes in the length of the hydrophobic region of the protein can increase or 
decrease the opening tension of the channel. For example, MscL channels might be engineered with 
an expanded hydrophobic region which matches PC14 bilayers. Our mismatch based 
theory would predict that the free energy versus acyl chain length curve would simply be translated
to higher $n$ so that the minimum $\Delta G_{0,\cal M}$ is realized for a PC14 bilayer. This shift should be 
measurable, reducing $\Delta G_0$ for PC16, PC18, and PC20 bilayers.
The reduction in mismatch may also allow MscL to be reconstituted into PC22 bilayers, allowing 
an additional data point. The proposed shift should also be measurable in EPR and SDSL 
measurements of residue proximity and mobility. The maximum mobility and separation should now be 
centered around $n=14$, perhaps permitting a clear measurement of the rise in induced tension for a PC10 bilayer 
predicted by the quadratic dependence of the line tension on the mismatch.

\subsection{Spontaneous curvature}
\label{sc}
Perozo {\it et al.} \cite{perozo1} have proposed that asymmetric bilayer stresses play a 
central role in MscL gating. They have proposed this model based on patch clamp, 
EPR, and SDSL experiments showing  that spontaneous curvature can induce MscL channel 
opening. Specifically, Perozo {\it et al.} find that MscL reconstituted into PC vesicles with high
enough concentrations of asymmetrically incorporated LPC stabilizes the open state of the channel, 
while MscL reconstituted into PC vesicles with symmetrically incorporated LPC, does not stabilize the 
open state. Unlike Keller {\it et al.} \cite{keller}, Perozo {\it et al.} 
have measured neither the spontaneous curvature for the mixed bilayer nor
the free energy difference between open and closed states as a function of LPC concentration.
In the absence of these quantitative experimental results, it is difficult to make concrete comparisons between 
our model and the experimental data. For large spontaneous curvature \cite{keller} but 
a relatively modest complementary mid-plane slope, the free energy difference between the two states due 
to spontaneous curvature is
\begin{equation}
\label{spont}
\Delta G_{C} \sim 2\pi (R_O-R_C) K_B CH' =
 -16 \left(\frac{20\ {\rm \AA}}{C^{-1}}\right)\left( \frac{H'}{-0.2}\right) \ kT,
\end{equation}
an energy typically large enough to stabilize the open state. 
In section \ref{ELB}, we have made some rather general arguments about the shape of the 
bilayer deformation energy landscape. We now return to this picture briefly to discuss the 
consequences of spontaneous curvature. In our discussion, we assumed that the line tension, 
$f$, was typically positive, but we remarked 
that this need not be the case in the presence of large spontaneous curvature.
If $f$ is negative, as depicted by the dotted curve in
figure \ref{FMR}, the only state stabilized by the bilayer is the open state, which very naturally 
gives rise to the open state stabilization observed by Perozo {\it et al.} \cite{perozo1}. 
Alternatively, this result can be understood from the predicted opening tension in equation \ref{opening_tension}.
When $\Delta G_0$ is bilayer deformation dominated, a negative $f$ implies that the opening tension is  
itself negative! A compressive force is required to stabilize the closed state. This
argument gives a tantalizingly simple explanation for the open state stabilization
but in the  absence of measured values for the spontaneous curvature induced by LPC,  
we can only conclude that spontaneous curvature could stabilize the open states for 
rather generic values of the parameters. An experimental consistency check of these results 
is fairly simple. Perozo {\it et al.} have incorporated LPC asymmetrically.  The same 
experiment might be repeated with $H_{II}$ phase inducing lipids which can also be used to 
generate spontaneous curvature but of the opposite sign. For the DOPC/DOPE system of Keller 
{\it et al.} \cite{keller}, the spontaneous curvature is known and tunable as a function of 
concentration. Our results predict that $\Delta G_0$ should be linear in $C$ \cite{ndan4}
and compatible with equation \ref{spont}.  

\subsection{Transition Rates}
\label{tr}
Although we have not discussed transition states in detail, the proposed model 
also predicts that the transition rates between states depends on the line tension, $f$. 
If we assume the transition rates to be given by the standard Arrhenius form:
\begin{equation}
k = k_0\exp( -G_B/kT ),
\end{equation}
where $G_B$ is the free energy barrier height for the transition and $k_0$ is roughly constant, 
the results can be understood intuitively by consulting figure \ref{FMR2}. At fixed tension, forward transition
rates (transitions to increasing radii) are reduced by line tension while backward rates are increased.
At any fixed tension, we would therefore expect the forward rates for MscL reconstituted into a PC16
bilayer to be faster than for a PC18 bilayer while the backward rates are slower since the PC18 bilayer has 
a greater line tension. The ratio of these rates can be calculated but depends on the radial location 
of the transition state. At the opening tension for each bilayer, it can be shown that if the 
bilayer deformation energy dominates, the predicted rates for the transitions $C\rightleftharpoons O$ will be faster 
for PC16 bilayer than for PC18 bilayer. Again, the ratio of these rates can be calculated in terms of the radii of 
the states and transition states. These two sets of qualitative rate predictions are straightforward 
to check experimentally.

\label{caveats}
In the argument above, we focused on the radial dependence of bilayer deformation free energy,
and fixed the other components of the state vector, $X$.
In principle, there is a potentially important piece of radial dependence we are missing. The
internal conformation may effectively couple the radius to the other parameters
in the state vector $X$, adding additional implicit radial dependence. For example, 
the thickness of the inclusion, $W$, is almost certainly a function of radius. 
It is also very natural to couple the mid-plane slope to the radius. 
We have ignored these dependences in order to develop an 
intuitive and simple one dimensional picture with as few undetermined constants and couplings as possible.
Provided that the bilayer deformation energy change is dominated by the 
radial change, this simplified model is a useful tool for understanding the bilayer-inclusion 
interaction. More elaborate models might easily be built from the general analytic framework
we have constructed. This framework will be described in a forthcoming paper. 

\section{Discussion}
 
We have argued that the bilayer deformation energy is harnessed by MscL to govern channel gating. 
Indeed, we have shown that a model which attributes the entire free energy difference between the
open and closed states to the bilayer deformation energy is compatible with the experimental data.
These results are somewhat surprising since it has been shown experimentally that the mutation of 
a single residue in the vicinity of the channel gate, can significantly effect channel gating 
\cite{ou}, \cite{yoshimura}. The protein conformational energy cannot be neglected in general. 
In fact, we have assumed that the protein 
conformational energy is large enough to constrain the channel geometry since we have assumed it is 
the bilayer which deforms rather than the protein. In principle, the closed state could have been 
stabilized by protein conformation alone, rather than mismatch, but exploiting bilayer deformation 
provides a robust mechanism for mechanotransduction, a design principle which functions in spite of 
the enormous number of nearly degenerate microstates endemic to proteins. Even for proteins as simple as 
myoglobin, Frauenfelder {\it et al.} \cite{frauen} have shown that the macroscopic conformation 
corresponds to an enormous number of structurally distinct microstates. These ideas have already been 
exploited for channel proteins. Goychuk and H\"anggi \cite{hanggi} have used this degenerate 
landscape to derive the empirical rate law for voltage gated channels. 
In light of these results, it is very natural to suppose that the protein conformational energy of 
the MscL protein gives rise to a vast number of nearly degenerate states as well. 
The bilayer deformation energy naturally breaks this degeneracy
and forms a mechanotransducing channel. The ensemble of microstates we observe as the closed states is 
stabilized at low applied tension by the line tension, while the ensemble of microstates we observe as the 
open state is stabilized by high applied tension. 
The importance of bilayer deformation in mechanotransduction may help to explain why there are no obvious 
sequence motifs associated with mechanosensitivity \cite{Bass} since a mismatch requirement does not imply 
sequence specificity. Harnessing bilayer elasticity does have one noted disadvantage. The gating of a channel 
will be affected 
by the membrane environment which surrounds it. This is precisely what the experiments of Perozo {\it et al.} 
\cite{perozo1} have shown. In realistic cell membranes there is an enormous diversity of
proteins and lipids which would imply that the free energy, and therefore opening tension, in these membranes 
would be heterogeneous. Sukharev and co workers \cite{suk3} have evidence for exactly this variability for 
MscL in giant spheroplasts. 

The correspondence between our simple theoretical 
model for the gating of the MscL channel and experiment is at least strongly suggestive that
this mechanism is exploited by the MscL channel. Our model can also naturally explain the 
stabilization of the open state by LPC \cite{perozo1} as well as the ephemeral nature of 
the substates \cite{suk1}. The tractable nature and simplicity of the model allow extensive 
analytic calculations to be made, which have in turn lead to numerous experimental predictions, 
discussed in  section \ref{mscl_g}. Specifically we have predicted (i) a shift in the 
curve relating the free energy difference and acyl chain length when the hydrophobic thickness of 
the channel is altered, (ii) the dependence of the free energy on spontaneous curvature and,
in particular, on the concentration of spontaneous curvature-inducing molecules, and (iii) 
the qualitative dependence of the transition rates on acyl chain length and spontaneous 
curvature.  

We have developed an extensive framework for studying bilayer-inclusion interactions in the 
MscL system. The model we have discussed here is the simplest implementation of these results 
and a more thorough description of the model will appear in a future paper. There are several very 
natural extensions to the current work. For example, we have focused here on the radial 
dependence only, but, as we have briefly alluded to in section \ref{caveats}, 
there are two additional geometric parameters which may also play important 
roles in the function of the MscL channel. More detailed measurements of the rates and free energies
of the various states will no doubt prove our simplified model incomplete and provide motivation and 
insight into a more detailed model of channel gating. 
The simplicity and generality of the competition between applied tension and line tension suggests that 
it may be a quite general phenomena for mechanotransduction. We hope to apply similar ideas to other
mechanosensitive systems.  More generally, we are also intrigued by the possibility of finding 
analogous bilayer deformation driven conformational changes for other transmembrane proteins which do 
not exhibit mechanosensitive function, perhaps illuminating a more general qualitative design 
principle for the function of transmembrane proteins.

\section{Methods}
\subsection{Lipids}
When discussing the lipids used by other authors, we have used the same naming convention they employed:
10:0 dicaproyl-phosphatidylcholine (PC10), 
12:0 dilauroyl-phosphatidylcholine (PC12),  
14:1 dimirstoyl-phosphatidylcholine (PC14),
16:1 dipalmitoleoyl-phosphatidylcholine (PC16), 
18:1 dioleoyl-phosphatidylcholine (PC18,DOPC), 
20:1 Eicossenoyl-phosphatidylcholine (PC20),
lysophospholipid (LPL), 
lysophosphatidylcholine (LPC),
dioleoyl-phosphatidylethanolamine (DOPE). 
We have preformed calculations based on elastic moduli measured by Rawicz {\it et al.} \cite{length}.
To extrapolate the effects of changes in the acyl chain length, we use the moduli for a typical 18:1 
phosopholipid (1,2-dioleoyl-sn-glycero-3-phosphocholine) and the scaling relations listed below
\begin{equation}
\begin{array}{cclcc}
 2a       & = & 37\ {\rm \AA}                            & \propto & a^1 \\
 K_A      & = & 0.64\  kT\, {\rm \AA}^{-2}               & \propto & a^1 \\
 K_B      & = & 21\  kT                                  & \propto & a^3 \\
 {\cal K} & = & 2.1 \times 10^{-3}\ kT\, {\rm \AA}^{-3}  & \propto & a^0
 \end{array} 
\end{equation}
where $2a$ is the equilibrium thickness of the bilayer, $K_A$ is the area expansion modulus, 
$K_B$ is the bending modulus, and 
\begin{equation}
\label{calk}
{\cal K} \equiv \sqrt{2}\left(\frac{K_A^3K_B}{a^6}\right)^{1/4},
\end{equation}
is the composite elastic modulus appearing in the line tension, $f_U$. The proportionality to 
bilayer thickness is approximate only and is deduced from treating the bilayer as a thin shell.
For more elaborate models and scaling arguments, see Rawicz {\it et al.} \cite{length}. By 
fitting the Peak-to-Peak head group thickness \cite{length} for saturated and mono-unsaturated lipids, 
we find that the relation between bilayer thickness and acyl chain length is roughly
\begin{equation}
2a = 1.3n+17\ {\rm \AA}.
\end{equation} 
Again, more elaborate models and scaling arguments are found in Rawicz {\it et al.} \cite{length}.

\subsection{Line tension}
In terms of $h(r)$, the bilayer mid-plane position, and $u(r)$, half the difference of the bilayer
thickness and the equilibrium thickness, 
the mean curvature contributions to the free energy density are \cite{helfrich} 
\begin{equation}
{\cal G}_{\rm B} = {\textstyle\frac{K_B}{2}}[\underbrace{\left(\nabla^2 h\right)^2+
\left(\nabla^2 u\right)^2}_{\cal M}-\underbrace{C_+\nabla^2[h+u]-C_-\nabla^2[u-h]}_{\partial \cal M}],
\end{equation}
where the variation of the ${\cal M}$ terms contribute to the action in the bulk (bilayer), 
the $\partial{\cal M}$ terms are total derivatives and can be evaluated at the
interface, the constant terms are dropped, $K_B$ is the bending modulus for the bilayer, 
and $C_\pm$ are the spontaneous curvatures of the upper and lower
leaflets, respectively. The tension contribution to the free energy density is \cite{goulian}
\begin{equation}
{\cal G}_{\alpha} = \underbrace{{\textstyle\frac{\alpha}{2}}\left(\nabla h\right)^2}_{\cal M},
\end{equation}
where the thickness deformation term can be ignored.
The interaction free energy density between the two layers is \cite{huang}
\begin{equation}
{\cal G}_{\rm I}= \underbrace{\frac{K_A}{2a^2}u^2}_{\cal M}, 
\end{equation} 
where $K_A$ is the area expansion modulus and $a$ is the equilibrium thickness of a leaflet.

The equilibrium equations that result from the minimization of $u(r)$ and $h(r)$ are
\begin{eqnarray}
0 = \frac{\delta G[u, h]}{\delta u} &=& \left[K_B \nabla^4+\frac{K_A}{a^2}\right]u(r), \\ 
0 = \frac{\delta G[u, h]}{\delta h} &=& \left[K_B \nabla^4-\alpha \nabla^2\right]h(r). 
\end{eqnarray}
In the asymptotic regime, large $R$, we can ignore the curvature of the interface and replace the Laplacians
with $d^2/dr^2$. For $u(r)$, the radially decreasing solutions are exponentials with complex wave numbers
\begin{equation}
\beta_\pm = -\left(\frac{K_A}{a^2K_B}\right)^{1/4}e^{\pm i\pi/4}.
\end{equation}
For $h(r)$, there is only a single radially decreasing solution with decay length:
\begin{equation}
\beta_h = \sqrt\frac{\alpha}{K_B}.
\end{equation}
At the interface, the bilayer deforms to match the hydrophobic thickness of the protein:
\begin{equation}
u(R) = U = {\textstyle\frac{1}{2}}W-a.
\end{equation}
If the inclusion interface is relatively flat and the top and bottom surfaces of the bilayer are normal
to the inclusion interface as depicted in figure \ref{basic}, the radial slope of $u(r)$ is zero at the interface:
\begin{equation}
u'(R) = 0,
\end{equation}
although this too may be used as a geometrical parameter \cite{ndan4}.
For $h(r)$, we set the radial slope at the interface:
\begin{equation}
h'(R) = H'.
\end{equation}  
At infinity, the bilayer is unperturbed, implying
\begin{eqnarray}
u(\infty) &=& 0, \\
u'(\infty) &=& 0,\\
h(\infty) &=& 0.
\end{eqnarray}
The equilibrium solutions can be found by matching the boundary conditions above.

The line tension is
\begin{equation}
f = \int_R^\infty dr \left({\cal G}_{\rm B}+{\cal G}_{\alpha}+{\cal G}_{\rm I}\right).
\end{equation}
Integrating this equation by parts twice yields a bulk term proportional to the 
equilibrium equations, which is zero since the equations are satisfied, and surface terms.  
The total line tension is:
\begin{equation}
f = \underbrace{{\textstyle \frac{1}{2}} {\cal K} U^2}_{f_U} + \underbrace{{\textstyle \frac{1}{2}} 
\sqrt{K_B\alpha} H'^2}_{f_H} + \underbrace{K_BCH'}_{f_C}, 
\end{equation}
where $\cal K$ has been defined in equation \ref{calk}, $C\equiv {\textstyle \frac{1}{2}}(C_+-C_-)$ is the 
composite spontaneous curvature, and $f_H$ is the mid-plane deformation contribution to the line tension 
which is typically small enough to ignore.

\section{Acknowledgments}
We are grateful to Doug Rees, Tom Powers, Jan\'e Kondev, Klaus Schulten, Evan Evans, 
Sergei Sukharev, Eduardo Perozo, Olaf Andersen, Sylvio May, Ben Freund, and Mandar Inamdar 
for useful discussions, suggestions, and corrections. We also acknowledge support from
the Keck Foundation, NSF and the NSF funded Center for Integrative Multiscale 
Modeling and Simulation. PAW acknowledges support through an NSF fellowship.

\end{document}